\documentclass{ecscw2007}
\usepackage[utf8]{inputenc}
\usepackage{ucs}
\usepackage{graphicx}
\usepackage{url}
\usepackage[hidelinks]{hyperref}
\usepackage{xcolor}
\usepackage{longtable} 
\usepackage{comment}
\usepackage{pdflscape}
\usepackage{caption}

\emergencystretch=\maxdimen
\begin{document}

\title{\Large{\textbf{Geofreebie: A Location-Based Freecycling App to Support Forced Migrant Resettlement}}}

\author{Lucas Braun and Auriol Degbelo and Christian Kray}
\affiliation{Institute for Geoinformatics, University of Münster}
\email{lucasalan@gmail.com$|$auriol.degbelo@uni-muenster.de$|$c.kray@uni-muenster.de}

\maketitle

\begin{abstract}
Germany has witnessed an influx of forced migrants in recent years. Promoting social interaction with the local community is key to supporting the resettlement of these newcomers. Location-based freecycling services present important benefits due to freecycling's potential to bolster social engagement and location-based services' ability to adapt to the user's context. Yet, their potential to support forced migrants’ resettlement is yet to be examined. We conducted needs assessment interviews with 11 participants in Münster, Germany. We analyzed the interview results to develop user requirements for location-based freecycling services. We then implemented a subset of the user requirements as a prototype mobile app called Geofreebie. The evaluation of the app with 22 participants showed that Geofreebie offered two key advantages for forced migrants’ resettlement: it increased the size of their social network, and created a sense of community on their side. These findings can benefit researchers and developers of location-based services to support forced migrant resettlement.

\end{abstract}

\vspace{-1cm}


\section{Introduction}
\label{cha:introduction}

Germany has witnessed an influx of forced migrants in recent years - 1.7 million asylum seekers between 2015 and 2019 according to the \citep{BAMF2020}. Facilitating the social integration of these migrants in their new society is vital to social cohesion. Promoting social interaction of the forced migrants with the local community during their resettlement is key in this context. As a guide from the Council of Europe puts it: \textit{``enabling diverse positive interactions builds belonging and cohesion''} \citep{orton2012building}. `Interaction' here refers to everyday processes by which migrants engage with each other and with receiving communities; `positive interaction' denotes processes that help build networks of mutually supportive relationships in a way that contributes to a more cohesive society \citep{orton2012building}.

Several studies have investigated how forced migrant resettlement can be eased by the use of novel technologies \citep{almohamed_vulnerability_2016,schreieck_supporting_2017,verbert_refugees_2016,brown_designing_2016}. \cite{alam_digital_2015} found that digital inclusion and social inclusion are linked issues for refugees. A study in New Zealand found that in order to promote the social inclusion of refugees, researchers and policymakers \textit{must} consider information and communication technologies (ICT) \citep{diaz_andrade_information_2016}. Mobile phones thus have a key role to play in the promotion of the social inclusion of forced migrants. 

Previous work to promote social inclusion for sustainable integration of forced migrants has relied on intercultural computer clubs and intercultural activities at school. The come\_IN approach \citep{Schubert2011,Yerousis2015,Weibert2010a} brings together people of different backgrounds and ages to work on joint projects, realize individual ideas at the computer, study or play. For instance, come\_IN projects connected German locals with participants with a Turkish migration background in \citep{Weibert2010a}. They also connected local Palestinian volunteers with an academic background, and children from refugee camps in \citep{Yerousis2015}. \cite{bustamante_duarte_participatory_2018} brought together young forced migrants and local students to ideate on urban services for other newcomers during a case study in Münster, Germany. Complementary to these initiatives, we propose a freecycling-based approach in this work. We argue that freecycling platforms are one kind of location-based service with a unique potential to foster social integration of forced migrants. Freecycling is the act of getting rid of something you do not need any more by giving it to someone else who does need it, without asking for a payment of any kind in return. 

Several features of freecycling make it particularly relevant to the promotion of `positive interaction' (as defined above) during their resettlement. First of all, freecycling platforms depend on knowing the users' locations in order to provide them with listings that can be picked up without excessive cost to the collector. Such co-located exchanges of material goods, both among forced migrants and between forced migrants and locals who are not forced migrants, could help \textcolor{black}{to partially address social isolation}. Indeed, location-based social networks facilitate arranging to meet in person and the exploration of one’s environment \citep{lee_location-based_2013}. Location-based context filtering makes services easier to use by reducing information overload and tailoring what is offered to the individual user. Tailoring to a local context may be particularly powerful when the goal is to create social contact, because the probability of social connection between two individuals decreases with the physical distance between them \citep{scellato_socio-spatial_2011}. In addition to its environmental benefits and appeal to people with limited resources, freecycling leads to trust-filled interactions outside of kin groups \citep{nelson_trash_2009}, encourages civic engagement \citep{nelson_downshifting_2007}, and blurs binary boundaries of needs and consumption \citep{eden_blurring_2017}.

These benefits have significant potential to support the resettlement of forced migrants but have yet to be researched as such. This article aims to investigate the potential of a freecycling-based approach for the promotion of social integration during forced migrant resettlement. \textcolor{black}{Recirculation of goods is at the core of freecycling. As a result, freecycling is a sharing economy activity in line with the categories of sharing economy undertakings introduced in \citep{Schor2014}. Germany is thus a good place for the current study because a survey \citep{TNSEmnid2015} has reported the willingness of consumers to share things they own with others. 88\% of the respondents reported their readiness to share things with others. Of these, 79\% were willing to share only with people they know and 9\% reported being willing to share with both acquaintances and people they do not know. These figures suggest that at least some individuals of the population would be willing to participate in freecycling with people they do not know.}

We used human-centered design principles to structure our exploration. Human-centered design focuses on improving ease of use and usefulness of services based on user input, and has been successfully applied to geospatial services \citep[e.g.,][]{Cay2019,roth_user-centered_2015,Perebner2019}. We applied the human-centered design process in three phases. In the first phase we performed a needs assessment study, carrying out interviews and extracting context and user requirements from the data. In the second phase we developed a prototype location-based freecycling service based on the needs identified. In phase three we carried out a user study to evaluate the prototype design.

The contributions of the article are threefold: (i) insights from a need assessment study, which identified and triangulated the needs of forced migrants, non-migrant locals, and moderators regarding a freecycling service; (ii) an open-source location-based freecycling service that addresses those needs; and (iii) lessons learned from a 2-week trial evaluating the service.


\section{Related Work}
\label{cha:background}
The design of a location-based freecycling service that supports the resettlement of forced migrants requires i) an evaluation of the opportunities offered by geospatial technologies in this area, ii) a consideration of the best practices and common concerns when developing ICT to address forced migrant challenges, and (ii) an understanding of the strengths and limitations of freecycling platforms for promoting positive social contact. Following is a short summary of past research on each of these three topics.

\subsection{Geospatial Technologies for Forced Migrants}
\label{sec:geospatial_technologies}
Location-based services have numerous application areas, with the more mature including tourism, transportation, healthcare, assistance, social networking, and gaming \citep[see][]{Huang2018,Raper2007a}. In the context of (forced) migration, relocation (i.e., moving from one country to another) is an obvious task supported by LBS, and understanding its spatial patterns falls within the purview of LBS research. A preliminary attempt to characterize transnational migration using location data from network operators was presented in \citep{ahas2018measuring}. Besides, relocation, resettlement is another area where LBS can be useful. For instance, they can help newcomers explore and acquire a mental map of the cities where they resettle. Or as mentioned above, they may be useful for social networking  \citep[a location-based freecycling platform is a purpose-driven location-based social network in the sense of][]{Roick2013}. 

An example of location-based service to support spatial knowledge acquisition during resettlement was provided in \citep{Kochar2017}. The author compared pedestrian navigation using a traditional map view to a map + augmented reality (AR) view, and reported that the map + AR view improved the recall of street names. Closely related, but focused more on navigational aspects during resettlement is the \textit{Lantern} system \citep{baranoff_lantern:_2015}. Lantern is a service that allows forced migrants in the United States to learn about their new environment by scanning strategically-placed NFC (Near Field Communication) stickers with their phones. Interacting with the Lantern service requires the migrants to wave their phones over NFC tags. They then get context-based guidance depending on their location. Our work has a similar objective to these two - facilitate resettlement - but a distinct focus on fostering social interaction through a freecycling approach.   


\subsection{Freecycling}
\label{sec:freecycling}
In 2003, Deron Beal coined the term ``freecycle,'' a blend between the words ``free'' and ``recycle,'' when he founded The Freecycle Network in Tucson, Arizona, in the United States\footnote{\url{https://www.freecycle.org/about/background} (accessed: June 07, 2020).}. Making things available for free via the internet, however, has been popular for much longer (think software, music, etc.) \citep{eden_blurring_2017}. There are now innumerable freecycling platforms all over the world. The Freecycle Network alone claims to have millions of members in more than 110 countries.

In Münster, the most notable freecycling platforms take the form of Facebook groups. The biggest is ``Verschenk's Münster,'' a giving-only platform with roughly 28,000 members as of this writing\footnote{\url{https://www.facebook.com/groups/473505729373257} (accessed: June 07, 2020).}. The second largest is ``Foodsharing Münster,'' a group focused specifically on avoiding food waste, which currently had just over 8,000 members at the time of this writing\footnote{\url{https://www.facebook.com/groups/607791439294335} (accessed: June 07, 2020).}. eBay Kleinanzeigen, a German subsidiary of the international eBay corporation, is a third popular platform for the free peer-to-peer exchange of goods\footnote{\url{https://themen.ebay-kleinanzeigen.de/ueber-uns/} (accessed: June 07, 2020).}.

\subsubsection{Value}

It has been shown that freecycling leads to trust-filled interactions between individuals with little in common other than their participation in a freecycling group \citep{nelson_trash_2009}. Freecycling also encourages civil engagement and members tend to be more politically engaged in society than the average citizen \citep{nelson_downshifting_2007}. Freecycling has also been shown to blur binary boundaries that are normally quite stark in such systems of exchange. Freecycling platforms bring together consumers and producers, givers and receivers, those who have resources and those who do not. Members can embody both roles at the same time. The platforms connect digital and material worlds, demonstrating the potential of virtual communities to have a real-world impact on people's lives. Finally, they unite mainstream and alternative cultures in one community \citep{eden_blurring_2017}. Therefore, participation in such communities could be of great value to the social integration of forced migrants. 

\subsubsection{Challenges} Organizational policies within current freecycling systems tend to tightly control the nature of exchanges due to an overemphasized focus on the environment. These policies contribute to the dominance of user communities by those seeking only ``green-washed convenience.'' Such user communities exhibit less altruism and solidarity when compared with other online groups \citep{aptekar_gifts_2016}. Freecycling communities are also not the altruistic ``gift economies'' that platform administrators often claim them to be, since most members expect some kind of generalized reciprocity, i.e. a reciprocal reward that may only come after a delay or from a different person than the recipient of the original ``gift''. Transactions that take place through such systems are a ``hybridized form of exchange'', having characteristics of both gift giving and trading \citep{arsel_hybrid_2011}. In order to create a real sense of solidarity (not charity) in generalized exchange platforms, the forging of a group identity is key \citep{willer_structure_2012}. Some groups forge such an identity through a common interest in sustainability, but that identity excludes people who could benefit from the service in practical ways, such as by acquiring things for free that they otherwise cannot afford \citep{aptekar_gifts_2016}. 

Our idea of a freecycling-based approach to support social integration of forced migrants in their new communities does not impose such restrictions regarding topical emphasis (i.e., environment) and/or group identity (i.e. sustainability). Instead our key premise is that enabling interactions between the forced migrants and residents in the local community can provide \textit{opportunities}, for both, to build relationships. These relationships, in turn, are needed to undermine the development of parallel societies.

\subsection{ICT for Forced Migrants}
\label{sec:ict}
ICT has been shown to help newcomers build social capital in their new city, to build trust by facilitating connections between newcomers and members of the host community, as well as by creating a sense of belonging in a group \citep{almohamed_vulnerability_2016}.

\subsubsection{Challenges}

The design of ICT for newcomers involves the unique challenge of balancing multiple cultural contexts in one system \citep{almohamed_designing_2016}. Designers must also consider users who have limited abilities in the local language and perhaps with literacy in general, limited internet access, a need to understand particularly complex compliance and geospatial information, a high need for reliability and timeliness, and limited experience with geospatial technologies \citep{bustamante_duarte_exploring_2018}.

\subsubsection{Best Practices}
The following services were all developed to study different ways that ICT can ease the resettlement of forced migrants. In each case, the researchers identified some useful best practices when designing tech for newcomers.

\textit{Rivrtran} \citep{brown_designing_2016} is a mobile app, also from the United States, that helps forced migrants to communicate with people who don't share a common language by providing access to volunteer interpreters. The authors found the strengths of such a service were that it facilitated communication between refugees and people who could provide support, simultaneously supporting and encouraging the users to eventually get by without such a service, and made use of a voice-based user interface to accommodate users with limited literacy. They also found a major benefit of the app was that it supported forced migrants to build social capital simply by engaging in everyday conversation.

\textit{Tarjimly} is a similar service developed in the United States by Atif Javed, Aziz Alghunaim, and Abubakar Abid \citep{utley_how_2017}. It is a Facebook Messenger bot that connects refugees needing translation services with volunteers able to help in the moment. The service shows what happens when you give people a way to support forced migrants without being majorly inconvenienced: thousands of volunteers offer interpretation services that would normally cost a fortune. The creators chose to use Facebook Messenger as their platform because it is a tool that is already familiar to so many users.

\textit{Integreat} \citep{schreieck_supporting_2017} is a service designed to facilitate information dissemination to forced migrants in Germany. The authors outline a number of best practices when designing software for maximum information transmission in an intercultural context. While the focus of this paper is not on information transmission, Integreat's design principles are also useful when applied to user interfaces intended for other purposes. For example, icons should always be accompanied by text and services should be usable offline.

\textit{Moin} \citep{verbert_refugees_2016} is a gamified informal learning app, also from Germany. The authors highlight the value of using a human-centered design process when developing technology for use by forced migrants, and stress the importance of accommodating low language abilities. They also point out how usability depends not only on the design of the system but also the context of the user, and how the contextual differences of forced migrants and German locals meant the app was much less usable for forced migrants.

\citet{bustamante_duarte_exploring_2018} worked with forced migrants and social workers in Münster, Germany, reviewing 36 apps and services and identifying several best practices and gaps. They observed a lack of services focused on non-Arabic-speakers and supporting offline use. They highlighted the importance of supporting multi-directional exchanges, user collaboration, flexible visualizations, and geovisualizations. They also recommended the use of open-source data platforms to create systems where knowledge can be built up over time by numerous contributors. Finally, they suggested the potential value of better geospatial technologies and location-based features, which up until now have been largely absent from ICT designed for forced migrants.

\subsection{Summary}

In this section, we have attempted to summarize past research on social integration in the context of forced migration and the use of ICT to address this and other forced migrant challenges. We highlighted areas that need more research, including 1) the development of services that are explicitly designed to promote the social integration of forced migrants and 2) the adaptation of existing promising technologies – such as geospatial technologies, location-based services, and freecycling systems – to the unique needs of forced migrants. The method that guided our investigation is explained in the next section.

\section{Research method}
\label{cha:research_method}
The question explored in this work reads: ``how to support forced migrants' resettlement through a location-based freecycling service?" As said above, we used human-centered design principles to structure our exploration. \cite{Abras2004} indicated that there is a spectrum of ways in which users are involved during human-centered (a.k.a. user-centered) design, but the crux is that users are involved in some way. We applied the human-centered design process in three phases, with each phase answering our research question in a different way and producing a unique contribution. Phase one answered the question theoretically and produced a list of needs and user requirements. Phase two answered the question technologically and produced a prototype location-based service demonstrating the feasibility of the designs. Phase three answered the question in the field and resulted in validation of the design decisions and a preliminary assessment of the usefulness of the prototype.

Per the ISO definition \citep{noauthor_iso_2010}, human-centered design consists of 5 steps, 4 of which may need repeating several times. They are: 1) plan the human-centered design process, 2) understand and specify the context of use, 3) specify user requirements, 4) produce design solutions to meet user requirements, and 5) evaluate designs against requirements. Following is a step-by-step overview of how we applied these steps in our work. 

\textit{1. Planning the human-centered design process.} We outlined roughly when and how each of the four remaining steps would be achieved, leaving room for iteration. We identified three user groups to be at the center of the design process. In this initial phase, we also met personally with local experts from forced migrant and freecycling communities to glean their recommendations about how to implement our design process. Young forced migrants expressed interest in new technology and committed to help with the design process and find others to participate too. Freecycling system moderators told us how to get in touch with their users, shared common freecycling challenges and provided assurance that the design of a new freecycling platform for forced migrants would be supported by their communities.

\textit{2. Understanding and specifying the context of use.} In this study, understanding the context of use meant investigating why and how forced migrants currently make social contact in Münster. Additionally, it meant learning about existing freecycling systems in the city. We used context interviews to achieve both of these tasks. The details of this process are described in Section~\ref{cha:needs}.

\textit{3. Specifying user requirements.} In order to identify key user requirements for a location-based freecycling service, we began by extracting implied needs from the context interviews. These in turn were translated into user requirements. See Section~\ref{cha:needs} for the results of this process.

\textit{4. Producing design solutions to meet user requirements.} The second phase of the study was designing and implementing a prototype app. We decided to make a cross-platform mobile app based on the location-based service called LBS-Engine\footnote{\url{https://github.com/LEinfeldt/LBS-Engine}.}. See Section~\ref{cha:prototype} for a full description of the implementation process and results.


\textit{5. Evaluating the designs against the requirements.} The designs were evaluated in an iterative fashion, first by having individuals test the prototype in one of the researcher's presence, then by running a pilot deployment to a small group of fellow students, and finally by deploying the prototype app to a larger group for a two-week trial period. During the trial, we collected data from users as we had planned. See Section~\ref{cha:evaluation} for full details and a discussion of the results of the evaluation. 
 
\textcolor{black}{In line with the spirit of the design study methodology \citep{sedlmair2012design}, a reflection on the transferability of this work's results to other domains is presented in Section \ref{subsec:transferability}. \cite{sedlmair2012design} pointed out that design studies have three types of research contributions: characterization of a problem domain (e.g., through abstraction into tasks and data); a validated design; and reflection. This article offers the three types of contributions. The needs assessment study (Section \ref{cha:needs}) provides a characterization of the problem domain: how to connect (forced migrants); how to freecycle (local freecyclers); and how to moderate a freecycling platform (moderators). The prototype and its validation through user feedback is the subject of Sections \ref{cha:prototype} and \ref{cha:evaluation}. The reflection takes place in Section \ref{cha:discussion}.}

\section{Needs assessment study}
\label{cha:needs}
User modelling is an important part of LBS research \citep[as discussed e.g., in][]{Jiang2006}. \cite{Huang2018a} put user modelling into the three `core' of LBS (next to positioning and the communication of location information). \textcolor{black}{Thus, one objective of this first study was to assess the needs of forced migrants with regards to social interaction during resettlement.} A second objective of this study was to assess the needs of the users and moderators of existing freecycling platforms in Münster. An understanding of those needs is essential to the design of a useful and usable location-based freecycling service.


In preliminary discussions with experts on forced migrant resettlement and freecycling in Münster, we decided to focus on triangulating the needs of three user groups to develop a user-centered application. We used in-person interviews to understand and specify the context of use for each group. The first group was forced migrants living in the general vicinity of Münster. The second group was residents of Münster who participate in some capacity in one of the several freecycling communities in Münster. The third group was the moderators of said groups. Collecting input from these three groups at the same time is valuable because they provide three complementary perspectives to the topic of resettlement support using freecycling systems. Forced migrants know how to make social contacts as a newcomer. Freecyclers know how to freecycle. And freecycling moderators know how to make a freecycling service safe and easy to use.

\subsection{Participants}
We recruited forced migrant participants by approaching acquaintances that we met through mutual friends and by attending newcomer-friendly meet-ups (e.g., WelcomeC@fé Münster). We contacted the moderators of freecycling services through their publicly available online contact information and obtained their permission to recruit freecyclers by posting directly in the freecycling forums. Further participants joined the study through snowball sampling. For each user group, we \textcolor{black}{continued interviewing until participants' responses were no longer yielding significant new information about the focus topic}.

Five of the interviewees were forced migrants, three were members of existing freecycling platforms in Münster, and three were moderators of these platforms. Of the five forced migrant participants, two were between the ages of 18 and 25, and three between 26 and 35. Two migrated from Syria and the other three from Iran, Iraq, and Turkey. We interviewed four men and one woman. Regarding the freecycling platform members, 1 was between the ages of 18 and 25, and two between 26 and 35. Two were women and one was a man. Only one did not grow up in Germany but rather immigrated to Münster from Eastern Europe for university. The freecycling moderators represented the broadest age range, between 29 and 50 years old. They all came from Germany and were long-term residents of Münster. Again, two were women and one was a man.

\subsection{Procedure}
We performed a total of eleven context interviews over a period of 22 days in November 2018. The script we followed for each interview can be found in appendix A1. The interviews followed the semi-structured focused interview method. This allowed us to maintain non-directive management of the conversation and still gather very specific information about the topic in question \citep{flick_companion_2004}. The objective of the interviews was to understand the \textit{context} in which the user community performs the \textit{focus activity}.


The forced migrant interviews focused on making social contact as a newcomer in Münster. The freecycler interviews focused on freecycling and the moderator interviews focused on moderating freecycling. In the end, all user groups gave information about all three topics. Forced migrants and freecyclers talked about the important role of the moderators in their communities. Freecyclers and moderators talked about making friends through their freecycling systems. Moderators talked about their own freecycling practices and forced migrants described the importance of giving generously when making social contacts in a new place. This helped us identify the context of use of a service that both enables freecycling and promotes social interaction. 

Context of use is made up of three elements: user goals, user tasks, and user environment \citep{maguire_context_2001}. The interviews in this study were made up of five questions, each with its own purpose and each helping to gather information about some of the three context elements. The main guiding questions were supplemented with probing questions in case the user needed more prompting to cover the topic. The full sheet of guiding questions for each user group can be found in appendix A. All appendices are available as supplementary material, see Section \ref{sec:supplementary}.



\subsection{Analysis}
Learning about the context of use allowed us to identify the core tasks that would need to be performed in the system we were designing. The core tasks, in turn, served as a foundation for all of the subsequent elements of the design process:

\begin{center}
{\small
    context $\rightarrow$
    core tasks $\rightarrow$
    implied needs $\rightarrow$
    usage requirements $\rightarrow$
    design
}
\end{center}

In order to process the interviews, we partially transcribed the recordings into what are known as \textit{context scenarios}. Context scenarios are summaries of the interview responses, formulated in an easy-to-understand and vivid way. Each interview yields one context scenario, so there is overlap between the scenarios. Repeat information is consolidated in the next step. Context scenarios serve as the building blocks from which we derive both implied needs and usage requirements. As shown in Table~\ref{tab:context_analysis_elements}, context scenarios are made up of narrative statements with identifying user information (such as names) removed. An example context scenario can be found in appendix B. 

The next step is to break down the interviewee's narrative descriptions of their activities into their core tasks and other context elements. This coding method is similar to descriptive coding or process coding \citep{saldana_coding_2009}, but focuses specifically on ``tasks'' or actions taken by the interviewee\footnote{Descriptive coding summarizes in a word or short phrase – most often as a noun – the basic topic of a passage of qualitative data; process coding uses gerunds (“-ing” words) exclusively to connote action in the data \citep[see][]{saldana_coding_2009}.}. The core tasks are phrased as a short verb phrase in the present tense, such as ``plan spontaneous meet-ups,'' ``introduce yourself,'' or ``find out about open/safe social opportunities''. These were repeated quite often across interviews, allowing the final core task list to be consolidated quite a bit by removing duplicates.

\begin{table}[ht]
\centering

\begin{footnotesize}
\begin{tabular}{|l|p{6cm}|}
\hline
\textbf{Context analysis element} & \textbf{Structure}                                     \\ \hline
Context scenario                  & User \textless{}verb in present tense\textgreater{}... \\ \hline
Core tasks                        & \textless{}Verb phrase in present tense\textgreater{}  \\ \hline
Implied needs                     & In order to ... , ... must ... .                       \\ \hline
Usage requirements                & The user must be able to ... the system ...            \\ \hline
\end{tabular}
\caption{Context analysis elements and their structure}
\label{tab:context_analysis_elements}
\end{footnotesize}
\end{table}

The core tasks translated naturally into implied needs. Implied needs were determined by analyzing each core task and identifying the conditions that must be met in order for the task to be completable. These conditions were pulled directly from the context scenarios, so they can be linked back to the words of the interviewee. From implied needs to usage requirements is a small leap. The translation is mostly a grammatical shift to phrase the implied need in terms of a user and an unspecified system.

\subsection{Results}
\label{sec:needs-results}
We extracted a total of 114 context elements from the five interviews with forced migrants, 95 context elements from the three interviews with freecycling participants, and 114 context elements from the interviews with the freecycling moderators. Table \ref{tab:interviewtasks} presents the 20 most frequent tasks mentioned in the interviews by each of the user group, along with their respective frequencies. Since our primary interest in this work is the social integration of forced migrants, their needs are now discussed in detail. 

\subsubsection*{\textbf{Core tasks and implied needs}}
We identified six core tasks when making social contacts as a newcomer, and analyzed
the interviews to determine what forced migrants need in order to complete those
tasks. Below are the six core tasks, along with the most frequent implied needs (and the number of participants who pointed at them).

\begin{itemize}
    \item Core task 1: find time to be social
    \begin{itemize}
        \item Need: match their schedule with social opportunities (N=3)
        \item Need: maintain financial stability (N=2)
    \end{itemize}
    \item Core task 2: find potential contacts
    \begin{itemize}
        \item Need: meet an ``introducer'' (N=5)
        \item Need: be patient and open to chance (N=4)
    \end{itemize}
    \item Core task 3: find a reason to initiate contact with potential contact
    \begin{itemize}
        \item Need: find people with common interests (N=5)
        \item Need: share their talents and abilities (N=4)
    \end{itemize}
    \item Core task 4: communicate with potential contacts
    \begin{itemize}
        \item Need: have accommodation of low language abilities (N=5)
        \item Need: improve communication ability (N=3)
        \item Need: start conversation (N=3)
    \end{itemize}
    \item Core task 5: plan and execute meetups with potential contacts
    \begin{itemize}
        \item Need: invite others to their home (N=5)
        \item Need: plan spontaneous meetups (N=5)
    \end{itemize}
    \item Core task 6: avoid bad interactions with potential contacts
    \begin{itemize}
        \item Need: block people who are bothering them (N=2)
    \end{itemize}
\end{itemize}

Only the most frequent implied needs are mentioned here. The full list of implied needs can be found in appendix D.

\begin{landscape}
\begin{table}
\begin{footnotesize}
\begin{tabular}{|l|p{7cm}|l|p{7cm}|l|p{7cm}|l|}
\hline
   & \multicolumn{2}{l|}{\textit{\textbf{Forced migrants}}}                       & \multicolumn{2}{l|}{\textit{\textbf{Local freecyclers}}}                              & \multicolumn{2}{l|}{\textit{\textbf{Moderators}}}                       \\ \hline
   & Tasks                                                       & \% & Tasks                                                                & \% & Tasks                                                  & \% \\ \hline
1  & find people with common interests                           & 100            & choose whom to give offer                                            & 100            & approve people to join the group                       & 100            \\ \hline
2  & invite others to your home                                  & 100            & take a photo                                                         & 100            & contact people who are not following the rules         & 100            \\ \hline
3  & meet an “introducer”                                        & 100            & browse for interesting offers                                        & 67             & delete posts                                           & 100            \\ \hline
4  & plan spontaneous meetups                                    & 100            & contact the people who react the fastest                             & 67             & enforce the rules                                      & 100            \\ \hline
5  & be patient and open to chance                               & 80             & decide if it is worth meeting up                                     & 67             & resolve conflicts                                      & 100            \\ \hline
6  & share your talents and abilities                            & 80             & decide what item is not needed anymore                               & 67             & tell members the rules                                 & 100            \\ \hline
7  & find out about open/safe social opportunities               & 60             & decide who needs or will use the offer the most                      & 67             & ask questions to people before they can join the group & 67             \\ \hline
8  & go regularly to work or school                              & 60             & offer food                                                           & 67             & block people who are bothering you                     & 67             \\ \hline
9  & improve communication ability                               & 60             & post item                                                            & 67             & block users who are misusing the system                & 67             \\ \hline
10 & join a course or club                                       & 60             & say where to meet up                                                 & 67             & check for new member requests                          & 67             \\ \hline
11 & join organization that puts you in contact with many people & 60             & tell other users how you want to be contacted                        & 67             & check quality of offers                                & 67             \\ \hline
12 & learn about host culture                                    & 60             & announce who can have the offer publically                           & 33             & communicate with other moderators                      & 67             \\ \hline
13 & match schedule with social opportunities                    & 60             & be patient and open to chance                                        & 33             & delete all posts for one member                        & 67             \\ \hline
14 & solicit help                                                & 60             & check group regularly for something you need                         & 33             & delete outdated posts                                  & 67             \\ \hline
15 & start conversation                                          & 60             & clarify details of meet-up                                           & 33             & develop empathy for conflict parties                   & 67             \\ \hline
16 & add other as a friend on Facebook                           & 40             & complete the offer hand-over quickly                                 & 33             & field misuse reports and feedback                      & 67             \\ \hline
17 & ask others questions                                        & 40             & connect people with things to give away with people who can use them & 33             & identify old offers                                    & 67             \\ \hline
18 & ask people about social opportunities                       & 40             & describe the offer                                                   & 33             & look at every person who wants to join                 & 67             \\ \hline
19 & block people who are bothering you                          & 40             & evaluate potential contacts                                          & 33             & scan offers to find misuse                             & 67             \\ \hline
20 & connect with people from different generations              & 40             & explain need                                                         & 33             & take a step backward from conflicts                    & 67             \\ \hline
\end{tabular}

\caption{Example tasks most frequently mentioned in the interviews, along with their frequency (in \%). The topical emphasis of the interviews was tailored to the group of participants: how they connect (forced migrants); how to freecycle (local freecyclers); how to moderate a freecycling platform (moderators). The full list of tasks extracted from the data is available in the supplementary material.}
\label{tab:interviewtasks}
\end{footnotesize}
\end{table}

\end{landscape}

\subsubsection*{\textbf{User Requirements}}

Out of the context and needs we derived a list of user requirements for a location-based service to foster the social interaction of forced migrants. For example, the user must be able to:

\begin{itemize}
    \item see promotions of social opportunities in the system (core task 1)
    \item see who is an ``introducer'' in the system (core task 2)
    \item tell the system about food they have to offer (core task 3)
    \item see the interests of other users in the system (core task 3)
    \item tell the system about their talents and abilities (core task 3)
    \item choose the language of the system (core task 4)
    \item tell the system to say something to another user (core task 4)
    \item see the location of the homes of other users in the system (core task 5).
    \item tell the system to invite other users to their home (core task 5)
    \item tell the system to confirm or reject a suggested place to meet up (core task 5)
    \item tell the system that they are available to meet up spontaneously (core task 5)
    \item see information about other users in the system before contacting them (core task 6)
    \item tell the system to block another user (core task 6)
\end{itemize}

These too are exemplar requirements. The full list of user requirements can be found in appendix E.




\subsubsection*{\textbf{Shared context of use}}
As mentioned above, a context of use is made of three elements: user goals, user tasks and user environment. \cite{maguire_context_2001} further broke down `environment' into: physical environment where the product is used, technical environment (hardware and software), and social/organizational environment (e.g., attitudes and cultures). From the analysis, we noticed that all three user groups reported the following \textit{shared} context elements:

\begin{itemize}
    \item Goal: increasing personal happiness
    \item Task: invite others into your home
    \item Tool: smartphones (technical environment)
    \item Risk: uncertainty about strangers (social environment/attitude)
    \item Philosophy: patience and openness (social environment/attitude)
    
\end{itemize}

This shared context has important implications for location-based freecycling services aimed at fostering the social integration of forced migrants. In particular, the presence of a shared goal suggests the pertinence of investigating systems bringing together the three user groups more broadly beyond the LBS freecycling service that is the focus of this work. The presence of a shared risk, philosophy and task is evidence that shared practices can emerge from the interaction between the three groups. Finally, the presence of a smartphone as tool in the shared context suggests that the shared goal is best supported through an app (in lieu of a desktop application or a website for example).

\section{Prototype development}
\label{cha:prototype}
We came up with a number of design solutions to meet the above needs with features of a location-based freecycling service called ``Geofreebie''\footnote{\url{https://github.com/lbraun/geofreebie}.}. We did this by implementing a subset of the user requirements developed during the needs assessment phase. 


\subsection{Needs addressed}

Overall, we identified 55 forced migrant needs in regards to increasing social contact. Here we discuss five key needs which we tried to address with the prototype that we developed.\newline

\noindent \textbf{Contact with a local ``introducer''}: Similar to the findings of \cite{schreieck_supporting_2017} that a ``local contact person'' is the most important source of information for newcomers, we found that meeting an ``introducer'' is their most important source of social connection. All of the forced migrants with whom we talked explained the importance of meeting a local with a strong social network who is able to introduce newcomers to potential new contacts. Most described just one key person who had introduced them to many friends. \newline

\noindent \textbf{Home where contact can be developed}: Forced migrants commonly invite people into their homes as a strategy to make meaningful social contacts. All of our participants mentioned using this strategy at least once. To do this, however, forced migrants must have a home that can accommodate guests. They must be able to communicate their home's location to others and feel confident and safe doing so. \newline

\noindent \textbf{Accommodation of low language abilities}: 
Perhaps the most discussed topic in our interviews about making social contact in Münster was the challenge of connecting with people without knowing much German. All of our participants experienced and overcame this challenge. They highlighted the need both for translators and translated resources when forced migrants first arrive, but also for opportunities to learn the local language.\newline

\noindent \textbf{Basis to initiate contact}: A commonly known strategy for making social contacts, not just among forced migrants but also in many other circles of society, is to seek out people with common interests. All of our participants reported doing this. They explained commonalities provide a basis to initiate a first conversation and an excuse to meet up again.\newline

\noindent \textbf{Element of spontaneity and scheduling flexibility}: Newcomers have busy, shifting, and complex schedules due to the wide variety and irregularity of important tasks they must complete in order to settle into their new home \citep[p. 17]{bustamante_duarte_exploring_2018}. We heard repeatedly that in order to make social contacts, one must find opportunities that fit with one's schedule. Often this means being spontaneous and approaching strangers to join a football match, start a conversation in a public place, or to offer help.

\subsection{Technologies}
We chose to implement the service as a mobile app because forced migrants are more likely to use smartphones than other hardware \citep{xu_communication_2016}. All of our interviewees had smartphones and cited them as their main means of communicating and one of their main tools when making new social contacts. Because most people carry their phones with them throughout the day, mobile apps also open up possibilities for spontaneous social contact in the moment and the location where the context is most apt.
We used the open source LBS-Engine\footnote{\url{https://github.com/LEinfeldt/LBS-Engine}.}, an application template for the development of location-based services, as a foundation for our app. This sped up development by providing an initial framework for displaying a map, navigating via a tab bar and a sidebar, and presenting the user with various settings. The LBS-Engine uses the open source Cordova framework from Apache, which allows mobile apps to be developed using web app technologies like JavaScript, HTML, and CSS. A major benefit of using Cordova is that once the application is developed, it can be easily ported to several different operating systems without rewriting the code. Currently supported operating systems include Android, iOS, OS X and Windows. This makes services developed with Cordova more accessible because they do not depend on the user having one specific operating system. The LBS-Engine makes use of the open source JavaScript mapping package LeafletJS and freely available basemap data from OpenStreetMap. It also uses the React JavaScript library with mobile-friendly user interface elements provided by Onsen UI. We maintained these technology choices.

\subsection{Features}

Following is a list of some of the more interesting features we implemented. For each feature we describe why we found the feature was important for a location-based freecycling service and why we implemented it the way we did.

\subsubsection{Authentication}
We decided to use passwordless authentication to make it easy to sign up and sign in. During the needs assessment we heard that many in our user groups had no desire to create another account and remember yet another password. We chose to allow people to sign in with Facebook and Google because 1) Facebook is the primary platform for freecycling in Münster, meaning most freecyclers are already connected and familiar with it, and 2) among our participants, Facebook and Google/Gmail/YouTube were the two most commonly referenced digital services after WhatsApp, which doesn't provide its own authentication service. We decided to additionally give users the option to sign up with their email and a secure password. While having to enter these details is usually less convenient, several participants from our needs assessment study expressed a preference for services that did not require connection with large corporate social networks like Facebook. Providing this third option allowed users to make the choice themselves between convenience and anonymity.

\subsubsection{User approval}

Approving new users was identified as the primary task of freecycling moderators. Local freecyclers repeatedly mentioned feeling reassured about the trustworthiness of other users because of the approval process. We also learned that forced migrants have their own system for approving new social contacts, often employing the very same strategies as freecycling moderators. These strategies include asking simple questions to see if the other could provide ``normal'' responses, identifying if the other is a part of a group that the approver trusts, and keeping an eye out for conflicting political beliefs. After signing up, new users are redirected to a screen informing them that they will receive an email when their account is approved. Location-based services enable an additional means of approval. In the Geofreebie prototype, users are automatically approved to make an offer available based on their physical presence in Münster. A bounding box geofence surrounds the city limits and users outside of this fence are automatically shown as ``unavailable'' in the system. This reduces misuse of the system by filtering out users actually located in other cities, which is a common concern among freecycling moderators.



\subsubsection{User communication}

\begin{figure}[ht]
  \centering
  \includegraphics[scale=0.20]{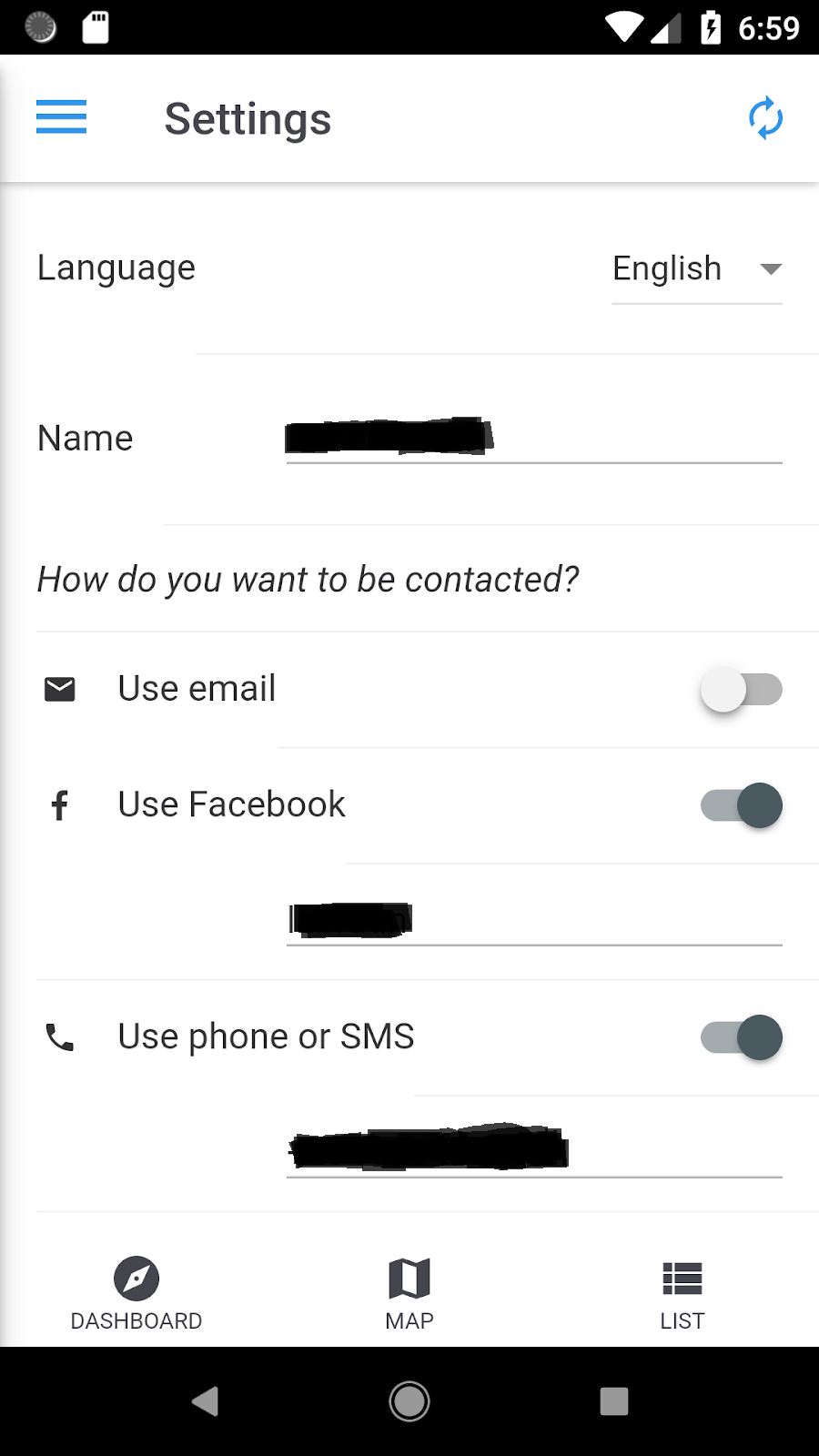}
  \includegraphics[scale=0.15]{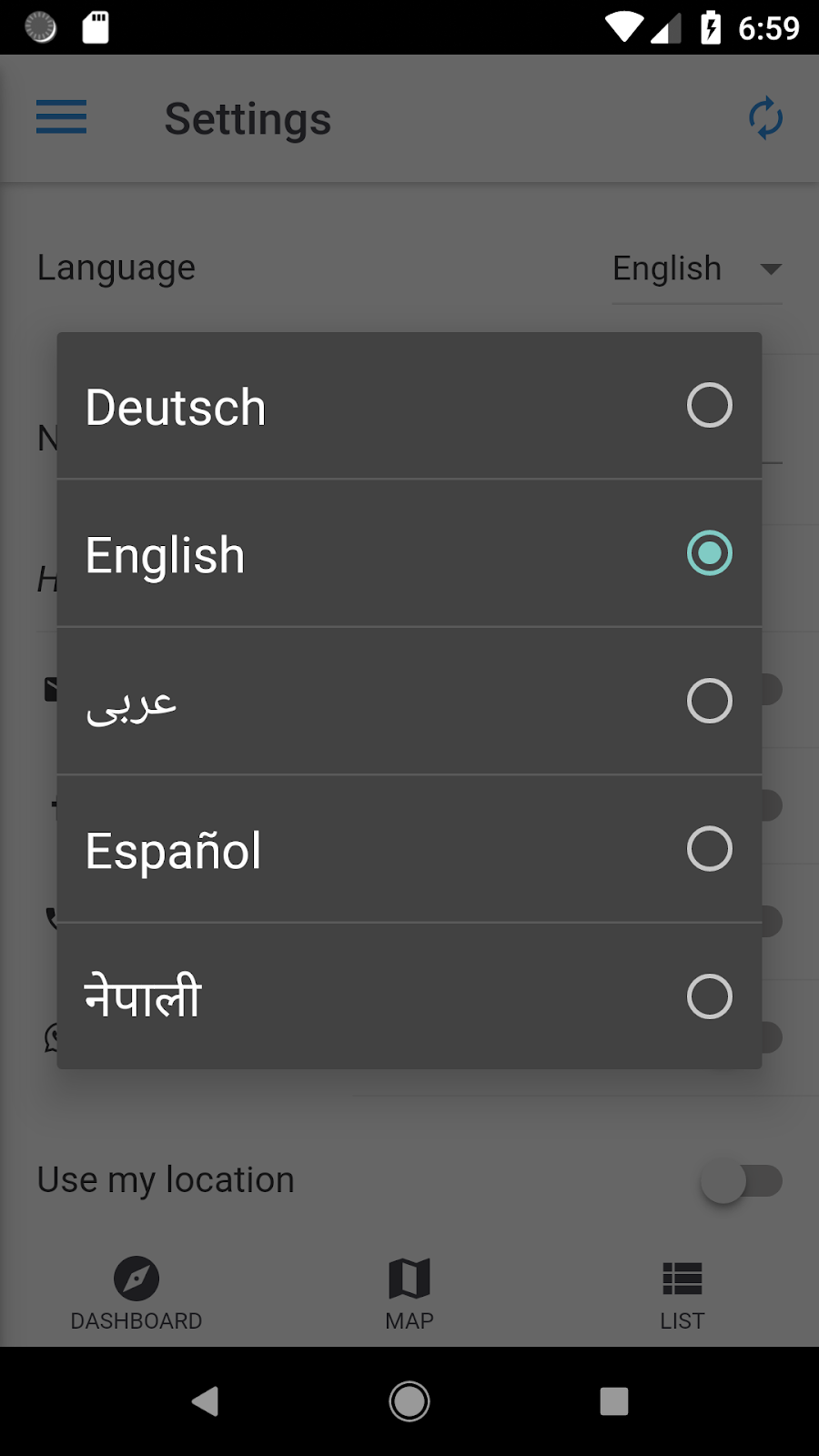} 
  \caption{Prototype settings pages}
  \label{fig:scr_settings}
  \label{fig:scr_locales}
\end{figure}

We designed interactions such that the person making the offer gets to choose the method of communication. This was reported to be common practice on Münster's Facebook-based freecycling platforms, where users would often state if they preferred to be contacted via comment or private message. Users choose how others can contact them on the settings page (see Figure~\ref{fig:scr_settings}, left). Toggle switches control their response to the question ``How do you want to be contacted?'' and when a certain contact mode is enabled, a text field appears where the contact details for that mode can be entered.

The four contact options we provided were email, Facebook, phone, and WhatsApp. These were the four most popular communication methods of the participants in our needs assessment study, matching the findings of \citet{xu_communication_2016}. After a user provides contact details, links are automatically generated from this information for other users to click on. Figure~\ref{fig:scr_map} shows an example of how these links appeared in the map view. 

\subsubsection{Language and localization}

We decided to make the app available in several languages in order to meet the need for accessibility to those with low German abilities. We implemented a language-switching feature called the locale menu, which is present in a number of places around the app (see Figure~\ref{fig:scr_locales}, right). We implemented the translation of the app's strings in one file (\textit{localizations.json}, see GitHub code). This made it simple for more languages to be added: a volunteer translator simply needed to go through and translate the list of 180 short strings, which usually took less than one hour, and then the new strings could be copied and pasted into a new element in the JSON array of localizations. By the time the trial had begun, the app had been translated from English into German, Arabic, Nepali, and Spanish. During the trial, another volunteer stepped forward to help with Amharic, and a Farsi native speaker expressed interest as well. The fact that the language menu is prominently present in several places (e.g., the help pages, the consent form, the settings page) and not just tucked away between other user preferences, is meant to accommodate those with low ability in one language while still encouraging them to learn other languages.


\subsubsection{Map and list views}

\begin{figure}[ht]
  \centering
  \includegraphics[scale=0.25]{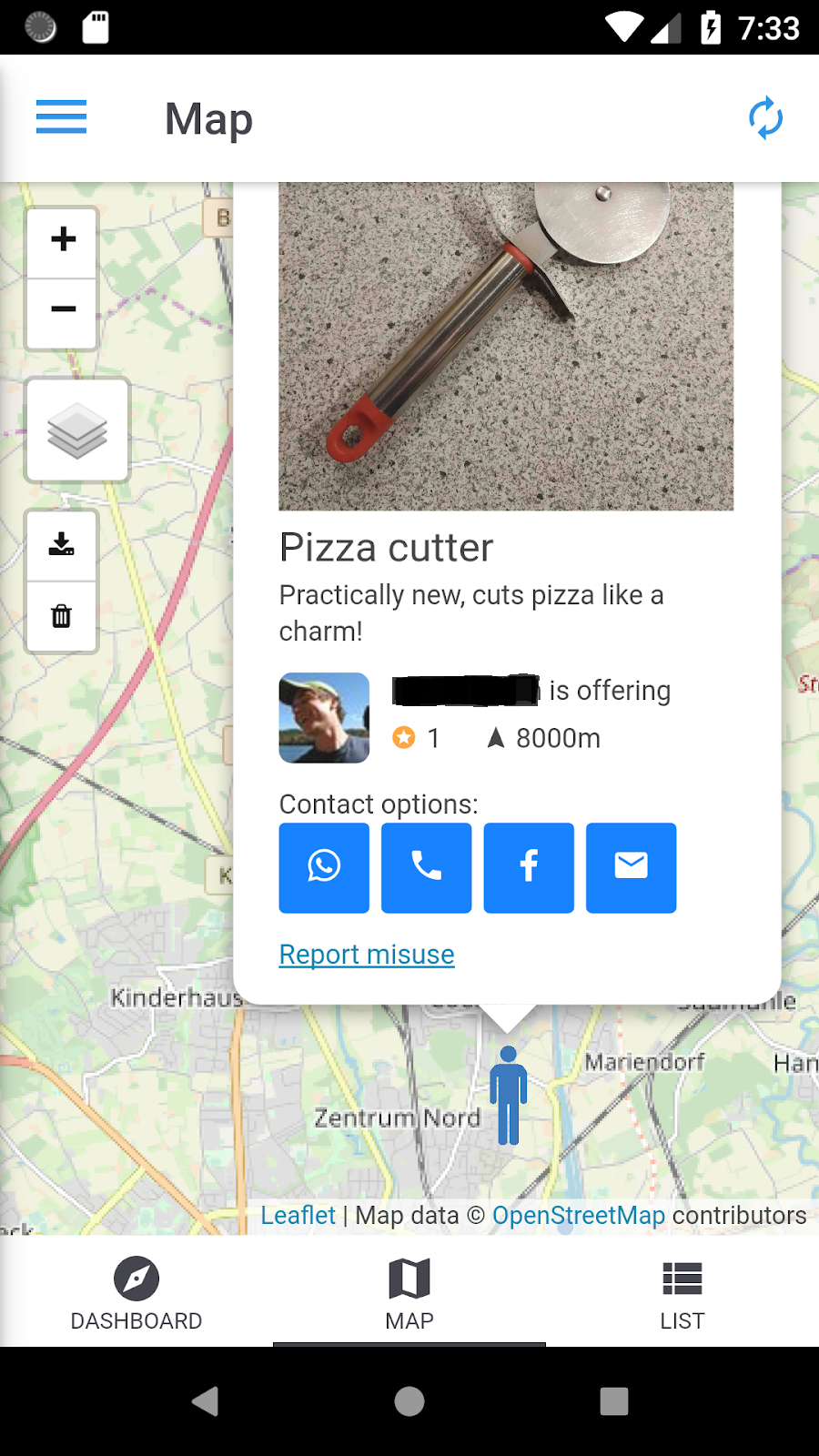}
  \includegraphics[scale=0.187]{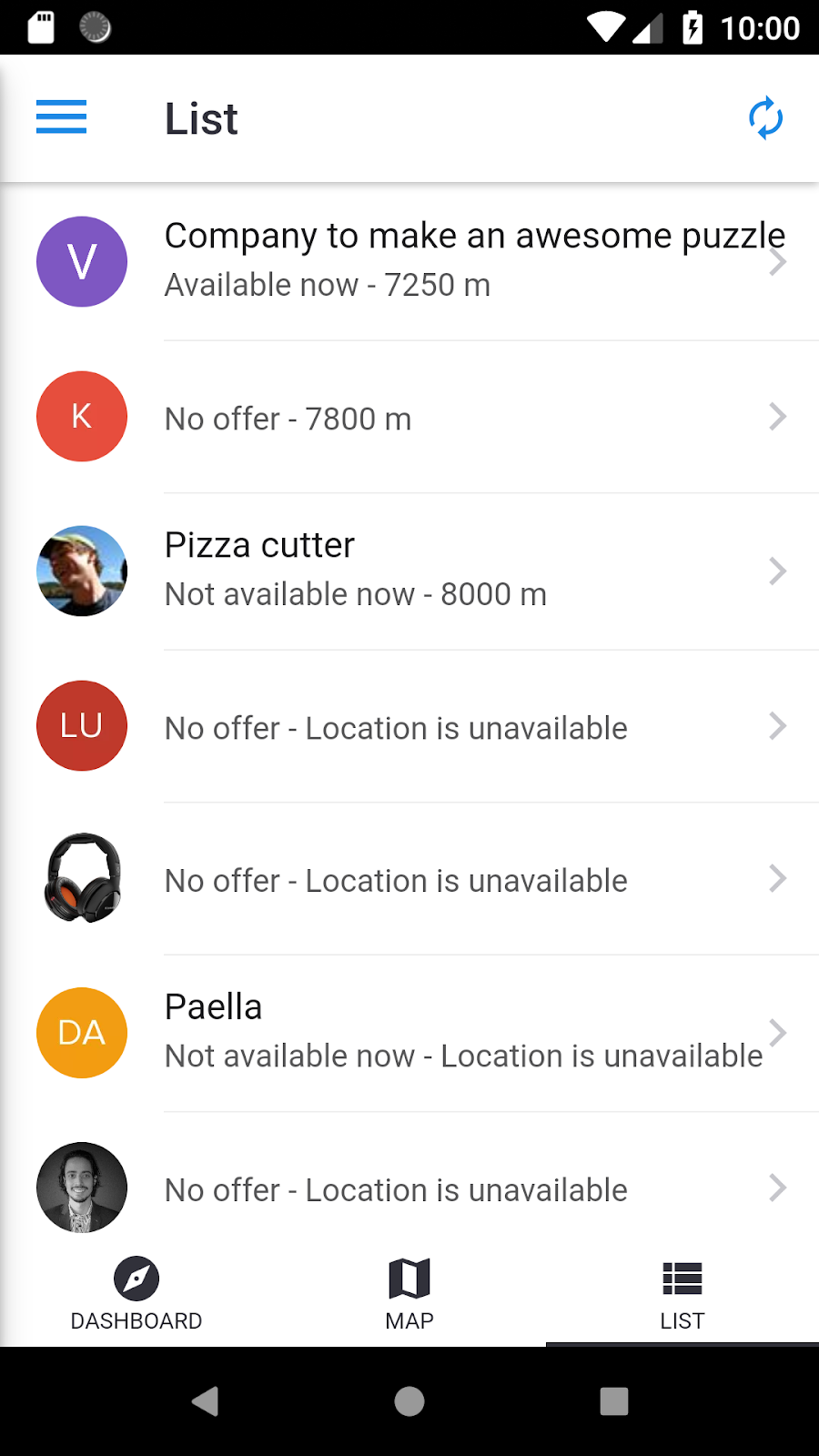}
  \caption{Prototype map and list views}
  \label{fig:scr_list}
  \label{fig:scr_map}
\end{figure}

The map and list views of the prototype were intended to meet several needs at once (see Figure~\ref{fig:scr_map}). The location-based features showing the distance to other users provides a basis to initiate contact (physical proximity) and facilitate those with limited language abilities to better visualize the offer and its location in the city. The Geofreebie star under the name of the person making the offer highlights the number of offers the user has already successfully delivered, thus indicating their status in the social network of the service and suggesting their potential as an ``introducer''. The app itself is also intended to play the role of the introducer by presenting potential contacts and suggesting reasons to connect.

\subsubsection{Reload button}

Pilot testing also revealed the need for a refresh button. Mobile users often do not trust their network connection and want to be sure that they have the latest version of shared data. Such a button gives them control to see the latest data when they want it, not at the standard interval when the app refreshes data automatically. We added a refresh button to the top right corner in the navigation bar, as can be seen in Figure~\ref{fig:scr_map}.

\subsubsection{In-app forms}

We chose to implement the study consent form and study surveys as in-app forms for a number of reasons. This allowed us to require users to fill out certain forms before using the system, ensuring that consent was timely. This also allowed us to recruit participants without meeting them in person, as no physical paperwork was required to participate in the study.

\subsubsection{Help}

Offering help and technical support is important in any digital service. In digital services for vulnerable populations, we found it becomes especially important. These services must not only explain the functionality of the app but also address concerns about safety and privacy, all in a way that is easily understandable by users with a diversity of language abilities. Our help page includes five sections: help, rules, contact, privacy, and consent.

\section{Prototype evaluation}
\label{cha:evaluation}
We conducted an evaluation study of the prototype app in February 2019 with 6 forced migrants and 16 local freecyclers. The objective of the trial was to evaluate the usability and usefulness of Geofreebie. The independent variable in the study was the user group of the participant (forced migrant or experienced freecycler). The dependent variables were the overall usability, the \textit{perceived} usefulness and the \textit{effective} usefulness of service. Usability was measured using SUS (the System Usability Scale, \cite{SUS,Lewis2018}, see appendix F4). Since we intended to evaluate the potential of the freecycling-based approach to foster social integration, and social interaction is \textit{a first step} towards integration, perceived usefulness touched on dimensions revolving around various aspects of social interaction. The questionnaire handed over to participants was loosely based on the indicators of social (dis)connectedness and perceived isolation defined by \citet{cornwell_measuring_2009}. Nine dimensions were probed: increased contact, increased contact with strangers, sense of solidarity created, sense of reliability created, sense of trust created, sense of community created, possibility for new friendships, increase in the size of the social network, and the reduced feeling of social isolation. See appendix F5 for the full list of questions. Finally, effective usefulness was assessed through the amount of goods exchanged during the study.  

\subsection{Participants}
We recruited forced migrant participants by contacting participants from our first study and by reaching out to other forced migrants we had met during the course of the project. We attended forced migrant meet-ups and helped interested attendees install and sign up for the service in person. We additionally shared a request for participants in a WhatsApp group for forced migrants and other members of a local club for newcomers to Münster. We recruited freecycling participants by posting in the two largest Münster freecycling platforms and by contacting participants from the first study who had expressed interest in later testing a prototype. All participants lived in Münster and were 18 years old or older. We did not control for specific age or experience. We motivated participation with a 10 Euro reward, delivered at the end of the study.

Of the 29 people who downloaded the app and created an account during the trial, we approved 25 for the study. From that pool, 22 consented to participate, filled out all of the surveys, and saw the trial through to the very end. Six of these self-identified as forced migrants. Of the six forced migrant participants, all were between the ages of 18 and 25. Four came from Syria, one from Turkey, and one form Eritrea. Four were men and two were women.

\subsection{Procedure}

The trial began on a Thursday and ended on a Wednesday in order to give participants two full weekends to try the app. Participants were required to use their own Android smartphone and internet connection. They installed the app via the Android Play Store. After installing the app and signing up for an account, participants were instructed to wait for their account to be approved. After their account was reviewed and approved, participants were sent an email to welcome them to the user community. The welcome email also contained links to instructions on how to get started with the app, the rules of the service, and information about the study. Participants were encouraged to reach out with any questions or concerns. Of the participants who showed interest in the study, four user accounts were not approved: two because the users signed up from outside of Germany, one due to insufficient information about the person, and one that was an existing user trying to sign in with a different email address. At the end of the trial, we contacted participants individually to arrange payment of the study reward and collect post-trial data. The app was removed from the Android Play Store and the back-end server was shut down to prevent continued use of the service. We had participants fill out the questionnaires to measure the app's usability and usefulness.



\subsubsection*{Consent and Surveys}

When participants logged into the app for the first time after approval, they were taken to the in-app consent form. This form asked users to agree to contribute their data to the study and separately requested permission to log particularly sensitive location data. Forced migrant participants received extra information about the consent to address ethical concerns when working with participants from vulnerable populations. After providing consent, participants filled out a demographic information questionnaire to give us a sense of gender, age, country-of-origin, and user group representation. They also completed the Lubben Social Network Scale (LSNS) six-item survey, a validated self-report measure of social isolation that correlates with negative health effects \citep{chang_validation_2018}. We used this survey to obtain a rough impression of the level of social engagement of our participants at the start of the study. The average LSNS score for the forced migrant group was 17.2 (out of a maximum of 30) while the average LSNS score from the other participants was 18.6/30, indicating moderate social engagement for both groups. The questions from these two surveys are listed in appendix F1 and appendix F2. Only after providing consent and completing the two surveys were participants granted full access to the app for the duration of the study.

\subsubsection*{Data collection during the trial}
\label{sec:datacollection}
Data was collected through both passive and active sampling. Passive sampling was conducted during the study by logging user updates to the system in the back-end server application. This was intended to help us understand how participants used the app. Active sampling was conducted through surveys that appeared to the user on the dashboard of the app as ``pending reviews''. The survey appeared immediately after completing an offer, the moment when freecyclers usually meet in person, and asked for a brief review of the hand-over. The intention of the active sampling was to collect data on any out-of-app interactions that occurred because of use of the freecycling system. We asked with whom, where, and how the contact was made. We also asked questions about satisfaction and likelihood of meeting up again in an attempt to measure the quality of the social contact. The exact questions of the survey can be found in appendix F3.



\subsection{Analysis}

All data collected in this third phase of the study was quantitative and required no qualitative analysis. We downloaded the data from the prototype services' database and used a Ruby script to process and summarize the results. We tallied the datapoints by action, by user, and by action and user. This allowed us to see how many people created offers, updated their profiles, adjusted the app settings, etc. The responses to the end-of-study SUS questionnaire were analyzed based on the classification of acceptable SUS scores developed by \citet{bangor_empirical_2008}. The responses to the questions about usefulness were considered individually.


\subsection{Results}

\subsubsection*{Passive Sampling Results}

Only 4 of the 23 participants posted offers in the app in total during the two-week trial, which is roughly 1 post for every 6 users. When compared with posting rates in other Münster freecycling platforms, this is abnormally frequent posting. One of the most active freecycling services in Münster had roughly 1 post for every 118 users in the same two-week period as the trial. One of the less active services saw fewer than 1 post per 1000 users. Thus, relative to the size of the community, the posting rate was very high, which indicates the potential to create many opportunities for social contact. It is also possible, however, that participants were motivated to post due to the novelty of the service and the desire to support the study. A longer trial would be required to see if the trend continued.

\subsubsection*{Usability Survey Results}
The average SUS score across all users was 82.6. Among forced migrant participants, the average score was 82.1 and among freecyclers it was 82.9.

\subsubsection*{Effective Usefulness Results}

Just one participant, a forced migrant, reported completing an offer. This interaction was reported to be arranged over WhatsApp with a user who was not a forced migrant and took place at the participant's home. The participant reported being satisfied with the interaction and believed they were likely to contact the person again.

\subsubsection*{Perceived Usefulness Survey Results}

\begin{figure*}[ht]
  \centering
  \includegraphics[scale=0.45]{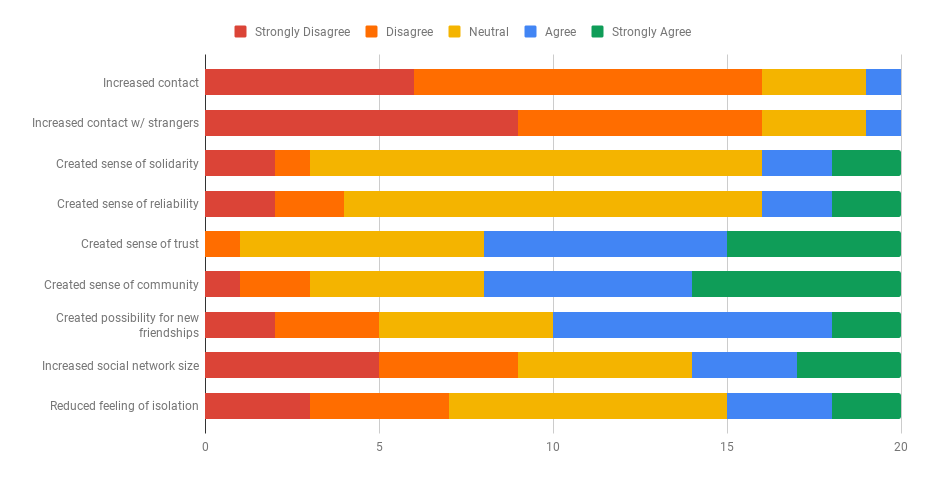}
  \caption{Perceived Usefulness survey responses: all responses}
  \label{fig:usefulness_survey_results}
\end{figure*}

In the final questionnaire, just one participant said the service led to increased contact with others during the two-week trial: the same participant who completed an offer. A majority of the participants in both user groups disagreed or strongly disagreed with the statements ``I think using this system increased my contact with others during the last two weeks'', ``I made contact with people outside of my normal circles through this system'', and ``I think this app increased the size of my social network in Münster.'' On the other hand, a majority of the participants in both user groups agreed or strongly agreed with the statements ``I feel like part of a community while using this system'' and ``I think the contacts made through this app are likely to lead to new friendships.'' Furthermore, a majority of both user groups responded neutrally to the statements ``I feel like I have a lot in common with other people using this system'' and ``I found I could rely on the other users of this system.'' The perceived usefulness survey results are summarized in Figure \ref{fig:usefulness_survey_results}.

\begin{figure*}[ht]
  \centering
  \includegraphics[scale=0.55]{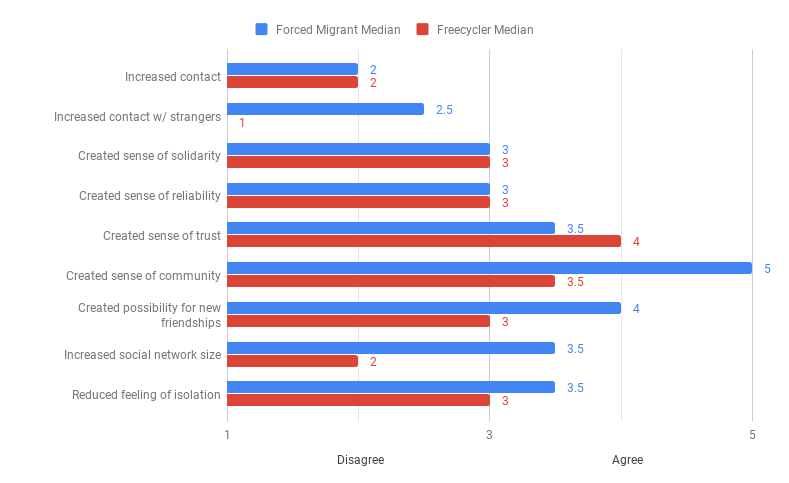}
  \caption{Perceived usefulness survey responses: user group comparison}
  \label{fig:usefulness_survey_medians}
\end{figure*}

The two user groups responded oppositely to just one question, the question about social isolation (see Figure \ref{fig:usefulness_survey_medians}). A majority of the forced migrant participants agreed with the statement ``I think this app made me feel less isolated from others in Münster'' while a majority of the freecyclers disagreed. 

\section{Discussion}
\label{cha:discussion}

\begin{quote}
When I fled from my homeland to Germany, my best friend was my cellphone. It really helped me and at the same time I helped many people because of my cellphone and my apps. I installed a navigator that worked offline on my phone and with maps from all the countries [that we were passing through]. I guided more than 100 people from various countries and the people trusted me because I already said [that] this [would be] happening like this and we are going to a place that is like this and when we got there [it was so].
\end{quote} 

This quote from a participant in our needs assessment study stresses the value of location information (and geospatial technologies) to forced migrants. Besides navigation during migration (scenario mentioned by the participant above), geospatial technologies can be useful for other stages of the migration process. Yet, there is a dearth of systematic assessments of this potential. The data collected and the location-based service presented in the previous sections can be useful to address this gap. This section now turns to the implications of the feedback provided by users both on the dimensions of usability and usefulness, compares features of Geofreebie with those of existing freecycling platforms, and points at limitations of the work.


\subsection{Usability}
A SUS score of 82 falls into the top quartile of the 206 studies analyzed in one meta study about SUS scores, and correlates with an adjective rating of ``good'' \citep{bangor_empirical_2008} and an ``A'' on the Sauro-Lewis curved grading scale \citep{Lewis2018}. Although a high SUS score does not guarantee acceptability in the field, it does suggest promise and a lack of major usability issues. More interestingly, both user groups gave close to the same average SUS score. This suggests that we successfully tailored the design solutions to both user groups through the human-centered design process.

\subsection{Usefulness}
Though the results of effective usefulness were promising, one offer completed is arguably too little to made definitive statements regarding this dimension of the evaluation. For this reason, we discuss here mainly the results of the perceived usefulness survey. We do not comment each and every dimension of social interaction probed, but focus only on the most significant observations.

\subsubsection*{Creation of new instances of social contact}

One participant reported having made a new social contact because of the app and described the interaction as positive and potentially leading to future repeat interactions. While this single result is not a significant indicator of the service's overall value in regards to generating new social contact between strangers, it does show the potential and feasibility of the idea. Participants generally disagreed with the statements ``I think using this system increased my contact with others during the last two weeks,'' ``I made contact with people outside of my normal circles through this system,'' and ``I think this app increased the size of my social network in Münster'' because they did not successfully complete any freecycling exchanges. Half of the forced migrants did think that the app increased the size of their social network, however. This may be due to the modern understanding that a social network can also be virtual. 

\subsubsection*{Creation of a trusting community}
Participants' general agreement with the statements ``I feel like part of a community while using this system'' and ``I think the contacts made through this app are likely to lead to new friendships'' also shows potential for reducing social isolation. Furthermore, only one participant disagreed with the statement ``I think I can trust the other users of this system.'' Feeling like part of a community, optimism about making new friendships, and contact with trustworthy people are all indicators of low social isolation \citep{cornwell_measuring_2009}.

\subsubsection*{Summary} Figure \ref{fig:usefulness_survey_results} illustrates that the freecyling-based approach has created a sense of trust and a sense of community in most participants, and as such fulfills \citet{orton2012building}'s desideratum of `positive interactions' between migrants and other residents in local communities. Even if the two groups provided occasionally diverging ratings regarding the gains, forced migrants seemed to have benefited the most. They had higher ratings on the ``increased contact with strangers", and ``increased social network size" dimensions. Thus, a freecyling-based approach is a promising way to foster social integration of forced migrants, next to approaches such as computer clubs \citep{Weibert2010a} or codesign workshops \citep{bustamante_duarte_participatory_2018}.

\subsection{Comparison with existing platforms}
While freecycling systems have great potential to reduce the social isolation of forced migrants, existing platforms have a number of problems in this regard. Moderators we interviewed openly shared how they applied their bias towards certain political groups and non-German speakers in their moderation tasks. The current dominance of Facebook as a platform for freecycling in Münster means forced migrants have to expose their identity in order to participate. Our platform gives the user full control over what elements of their identity they would like to share, thus reducing moderator and peer bias. Moderators also have less reason to be biased towards foreigners, because posts are geofenced, which limits the risk of spam from people outside of the geographic community. Another problem with existing platforms is that recognizing what offers are nearby requires a geographic knowledge of the city, knowledge that newcomers inherently lack. Unlike our platform, none of the existing platforms have integrated maps or location-based features beyond the scoping of offers by city. Furthermore, the current mono-lingual platforms alienate those who cannot understand posts and group rules. We designed our interface to be language and culture agnostic, allowing the user to see rules and interface elements in their own language and emphasizing pictures and location information over words of description in posts.

\subsection{Transferability}
\label{subsec:transferability}

\textcolor{black}{A closer, focused, look into user needs is in line with calls from previous work. Developing a `handbook of human needs' has been identified as one of the grand challenges of HCI research in \citep{Shneiderman2016}; and \citet{Talhouk2018} argued that ``Technological designs should \textit{reflect refugee needs}, experiences, and values" (emphasis added). Though the needs extracted from the interviews have helped to design a location-based service, they are relevant to Computer-Human interaction in general, and computer-supported cooperative work (CSCW) in particular. Freecycling platforms require deliberate coordinated actions and coordinative practices and hence are CSCW systems in the sense of \citep{Lee2015,Schmidt2011}. For instance, \cite{Lee2015} introduced seven dimensions of CSCW work, namely synchronicity, physical distribution, scale, number of communities of practice, nascence, planned permanence, and turnover. In regard to these dimensions, Geofreebie has the following properties: it supports asynchronous interactions between the participants (the interactions start asynchronously in virtual space, and may end up synchronously in physical space if two users decide to meet to exchange an artifact); coordinated actions may start with participants at completely different geographic locations and end with them being at the same geographic location to complete a transaction; the scale (i.e., number of participants) is potentially unlimited but was 22 during the evaluation; the number of communities of practice is at least three (i.e., freecyclers, moderators, forced migrants), but potentially unlimited as well; coordinated actions with Geofreebie were still in the nascence phase during the two-week trial; the planned permanence of the actions undertaken was arguably short-term given the two-week study; and the turnover was low during the evaluation of Geofreebie, probably due to the foreseeable time-frame of the experiment (the turnover of a fully-functional application may be high or low depending on the actual perceived benefit by the participants).}

\textcolor{black}{Previous work \citep[e.g.,][]{AbuJarour2017, bustamante_duarte_exploring_2018} has investigated challenges of forced migrants in Germany, and reported among other the limited proficiency in German/English as an important challenge. Our data reflects this as well (see `user characteristics' in the Supplementary Material), but goes beyond this. Interviews with our participants have provided evidence that any CSCW system for freecycling with forced migrants in Germany should support at least five elements (Section \ref{sec:needs-results}): the shared goal of increasing personal happiness, the shared task of inviting others into one's home, the shared tool of smartphones, the shared risk of uncertainty about strangers, and the shared philosophy of patience and openness. If these five elements are useful per se, some provide additional implications for existing research. For instance, the shared goal of increasing happiness suggests that \cite{Desmet2012}'s approach of possibility-design could be valuable in studying CSCW systems for freecycling with forced migrants. \cite{Desmet2012} proposed possibility-driven design as an approach that focuses directly on what makes people happy instead of removing prevailing problems. It breaks free from prevailing issues and channels creativity into the question: what are desirable future states? The shared context elements identified in Section \ref{sec:needs-results} provide a starting point for further research along these lines. Also, the shared task of inviting others into one's home makes the case for the integration of location-based information into these CSCW systems, while the shared risk raises a question, namely: how to increase participants' (perceived) safety during the use of these systems? There is a need for research building and deploying CSCW systems to systematically investigate these questions.} 



\textcolor{black}{Finally, another important lesson was that recruiting participants from a Facebook group can be challenging and unfruitful. For example, we posted one request for interviews in the largest and most active freecycling group, with more than 27,900 members and on average 17 posts per day. Only 10 members responded to this call and all but 1 of those backed out before meeting up to talk. These figures are mentioned to provide a basis for comparison with similar research efforts across platforms and countries.}


\subsection{Limitations}
Our initial two-week trial of the prototype was not long enough to evaluate the full usefulness of the service. The relatively short duration of the trial was limiting for three reasons. First, building up a large user community takes time. Freecycling systems work best when there are a lot of users. The more users that are actively posting and browsing for offers, the higher the chance that someone will offer something that another user actually needs. Second, as we identified in our needs assessment, both freecycling and making social contacts as a newcomer require good luck and patience. The longer someone is a member of a freecycling system, the more likely they are to find something they would like to offer or see an offer of interest. Third, building relationships takes time. Even when people contact someone through a freecycling service, it takes time to arrange an in-person meetup. If this meetup is the first of repeated meetups, it may be a week or more before the next meetup takes place. In our model where the system is the ``introducer,'' extra time is also required for the users to build a relationship of trust with the system.

\textcolor{black}{Besides, our recruitment strategies were biased towards forced migrant participants who were already social enough to be out meeting strangers, meaning less social people were underrepresented. Women and older forced migrants were also underrepresented, so our results may not fully reflect their needs and contexts. In addition, the small size of the sample does not allow generalization to all forced migrants, freecyclers, and moderators in Germany. Finally, though the implementation of the prototype has focused primarily on needs related to the creation of social contact and the accommodation of low language abilities, it is acknowledged that data privacy aspects are important for the successful adoption of Geofreebie in the long term. Exploring means to best address location privacy (e.g., how to implement ephemerality features discussed in \citep{Stroeken2015,ataei2017ephemerality}, or give users control over their data through the user interface elements suggested in \citep{ataei2018privacy}) provides nice avenues for future work.}

\section{Conclusion}
\label{cha:conclusion}
This work has assessed the potential of a freecycling-based approach for the promotion of social integration during forced migrant resettlement. We identified the context, core tasks, and specific needs of this group in regards to creating social contacts in their new city. We developed design solutions for a location-based service that supports forced migrant resettlement by helping them build their social network. We implemented some of these solutions and published them as an open-source mobile app to allow adaptation and continued development. A 2-week trial evaluating the app has shown the potential of the approach for positive social interactions in a real-world scenario.

\textcolor{black}{The needs of forced migrants collected through the interviews provide a basis for evaluating future systems supporting their resettlement. The triangulation of these needs with characteristics of stakeholders with different profiles (freecylers and moderators) is a unique feature of our work, and the shared context elements extracted provide a basis for the design of computer-supported cooperative systems involving the three types of users. Finally, supporting resettlement has primarily been done through intercultural computer clubs or intercultural workshops in the literature (e.g., the come\_IN  approach or the organization of co-design workshops bringing together forced migrants and locals at school). The positive feedback from forced migrants on the usefulness of our prototype suggests that freecyling-based LBS systems are a promising approach as well.}

There is great opportunity to continue the work started in the research conducted for this article. First, many of the user requirements developed in the needs assessment phase of this work remain unimplemented in the prototype we built, and these could be added to the next version of the Geofreebie app.  Second, the prototype would benefit from the further evaluation on a larger scale. This evaluation could include not only more participants, but also with more time and perhaps with trials in other cities, to further understand the value of a freecycling-based approach for social integration in the long term. 

\section*{Acknowledgements}
The authors gratefully acknowledge funding from the European Commission through the Erasmus Mundus Master in Geospatial Technologies (Erasmus+/Erasmus Mundus program, grant agreement FPA-2012-0191, \url{http://mastergeotech.info/}) and the GEO-C project (H2020-MSCA-ITN-2014, Grant Agreement Number 642332, \url{http://www.geo-c.eu/}).

\section*{Supplementary material}
\label{sec:supplementary}
The supplementary material including all appendices is available at \url{https://doi.org/10.6084/m9.figshare.13530689}.

\bibliographystyle{ecscw2007}
\bibliography{0_JLBS2020}

\end{document}